\documentclass[aps, pre, reprint, superscriptaddress, hyperref, twocolumn]{revtex4-1}
\usepackage{graphicx} 
\usepackage{dcolumn} 
\usepackage{bm} 
\usepackage{hyperref} 
\usepackage{longtable} 
\usepackage[T1]{fontenc}
\usepackage{times}
\usepackage{xcolor}
\usepackage{wrapfig}
\usepackage{float}

\graphicspath{{./figures/}}

\renewcommand{\BibitemShut}[1]{}

\begin{document}

\title{Mapping Disorder in Entropically Ordered Crystals}

\author{James \surname{Antonaglia}}
\affiliation{Department of Physics, University of Michigan, Ann Arbor, MI 48109, USA}
\author{Greg \surname{van Anders}}
\affiliation{Department of Physics, University of Michigan, Ann Arbor, MI 48109, USA}
\author{Sharon C.\ \surname{Glotzer}}
\affiliation{Department of Physics, University of Michigan, Ann Arbor, MI 48109, USA}
\affiliation{Department of Chemical Engineering, University of Michigan, Ann Arbor, MI 48109, USA}
\affiliation{Department of Materials Science and Engineering, University of Michigan, Ann Arbor, MI 48109, USA}
\affiliation{Biointerfaces Institute, University of Michigan, Ann Arbor MI 48109, USA}

\date{\today}

\begin{abstract}

Systems of hard shapes crystallize due to entropy. How is entropy distributed
among translational and rotational microscopic contributions?  We answer this
question by decomposing thermal fluctuation of crystals of hard hexagons into
collective modes, a generalization and quantification of the Onsager picture of
hard rod liquid crystals. We show that at densities both near densest packing
and near the solid-hexatic melting transition, solids of hard regular hexagons
hold most of their entropy in translational degrees of freedom.

\end{abstract}

\maketitle

Plato's order/chaos dichotomy~\cite{timaeus} has a long intellectual history
that achieves its most precise mathematical formulation in the theory of phase
transitions~\cite{landauphase}. However, phase transitions can be driven by
entropy~\cite{Onsager1949,Alder1957,wood}, which is historically associated with
disorder~\cite{clausiuswarmtheorie}. Entropy-driven phase transitions confront
Plato's dichotomy with an apparent paradox: a macroscopically ordered system can
exhibit more microscopic disorder than a macroscopically disordered
system~\cite{frenkel,ordviaent}. Entropy-driven liquid
crystalline~\cite{frenkel} and crystalline order~\cite{entint} can be
rationalized through intuitive arguments for simple enough model systems, but
resolving the paradox for more complex types of order~\cite{amirnature,
Damasceno2012, escobedo, geissleryang} requires a detailed description of the
entropy in an ordered system. Quantifying entropy contributions is
complicated by the fact that entropy arises from statistical considerations, and
is inherently macroscopic unlike energy, which trivially decomposes into a sum
of microscopic contributions.

Here, we quantify contributions to the entropy of hard particle crystals by
decomposing thermal fluctuations in two-dimensional systems of hard regular
polygons into collective modes. Collective vibrational (phonon) modes have been
studied previously for hard spheres~\cite{Stillinger1967,Honda1976}, and
propagating rotational (libron) modes have been studied in detail in molecular
crystals~\cite{Schomaker1968,Venkataraman1970,Raich1971}. We compute the
dispersion relations for phonon and libron modes for hard
hexagons~\cite{anderson2017shape}, and show that the macroscopic system entropy
is partitioned into phonon and libron contributions. From these contributions,
we construct a ``map,'' by wavevector, of the contributions of microscopic order
that lead to macroscopic order. For macroscopic order in systems at densities
just above crystallization and close to maximal packing, we find a
predominance of translational disorder. We find that the reason for the
predominance of translational order, however, is different at low and high
densities. Moreover, we find that between density extremes systems receive
primary contributions from both translational and orientational disorder at
different wavevectors. Our results give a concrete picture of ``disorder within
order,'' and offer an intuitive explanation of recent, counterintuitive findings
on the behavior of densely packed anisotropic colloids.\cite{packingassembly}

\begin{figure}
  \centering
  \includegraphics[width=0.5\textwidth]{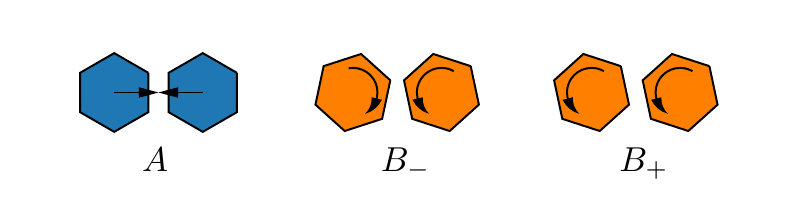}
  \caption{Microscopic degrees of freedom which contribute to collective modes. Below each image is listed the parameter names used for the second derivatives of $\mathcal U_{1,2}$, the effective interparticle potential, with respect to that mode. The first and second relative displacements correspond to stiffnesses that penalize gradients in both crystal displacements and body orientation, respectively. The third type of displacement corresponds to a stiffness that penalizes homogeneous or cooperative rotation and gives rise to a libron dispersion gap. There is no corresponding cooperative translation term because of global translational symmetry.}
  \label{force_diagram}
\end{figure}

\begin{figure}
  \centering
  \includegraphics[width=0.5\textwidth]{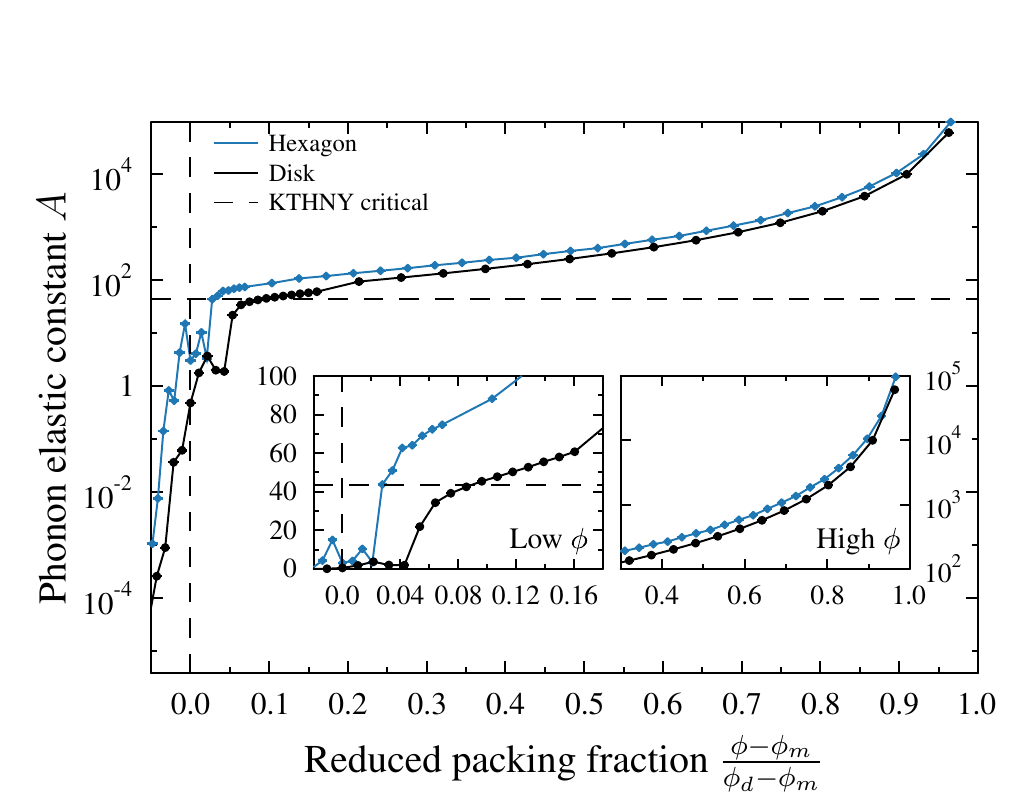}
  \caption{Phonon elastic constant for crystals of hard disks (black) and hexagons (blue) at all densities, scaled to range from 0 (melting) to 1 (densest packing). The black dashed line indicates the critical value for $A$ at the hexatic-solid transition, predicted by the KTHNY melting theory. Both solids exhibit similar stiffness constants at the same normalized densities at high density. Insets are magnifications in the vicinity of the melting transition and densest packing, respectively. Error bars indicating 1-$\sigma$ standard deviations of elastic constants from best fits of four independent simulations to theoretical dispersion curves are smaller than the plot markers.}
  \label{couplings}
\end{figure}

\begin{figure*}
  \centering
  \includegraphics[width=\textwidth]{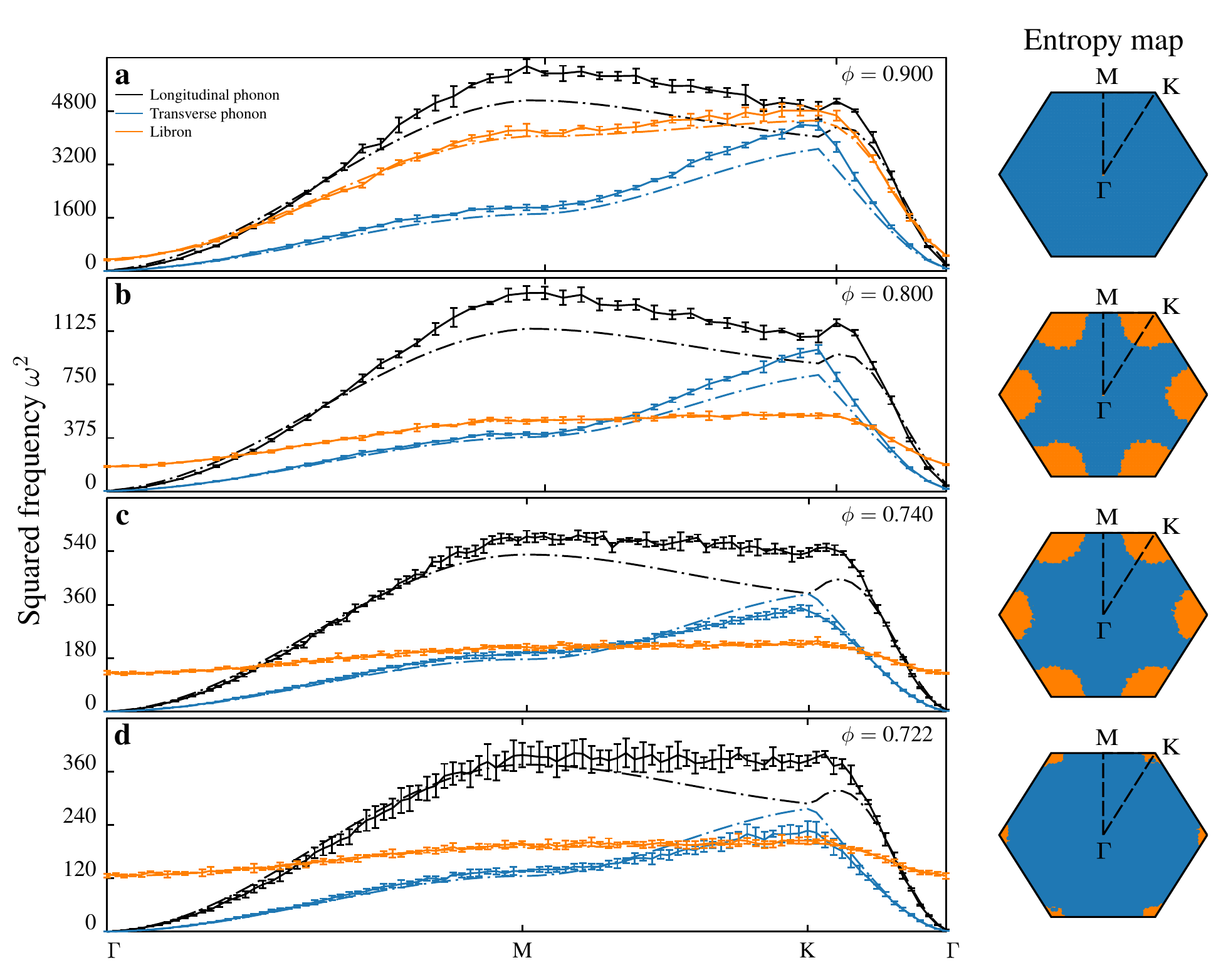}
  \caption{Dispersion relation for hard hexagon crystals at various packing fractions. All dispersion relations decrease with decreasing density, but the phonon dispersions become ill-defined at densities close to the hexatic-solid phase transition. Right panels are the first Brillouin zone colored corresponding to the lowest frequency mode at each $\vec k$-vector; orange indicates the libron mode and blue indicates the transverse phonon mode. Broken lines indicate theoretical dispersion relations with best-fit stiffness parameters. Error bars are 1-$\sigma$ sample standard deviations of the dispersion frequencies averaged from four independent simulations.}
  \label{dispersion}
\end{figure*}

Dispersion relations relate the frequency and wavelength of vibrational and
librational modes in crystals. We computed dispersion relations for phonons
and librons in a crystal of hard hexagons from NVT simulation using the
hard-particle Monte Carlo (HPMC) plugin~\cite{Anderson2016} for
HOOMD-blue~\cite{Anderson2008b, Glaser2015}. We use systems of $N=10,000$
hexagons near the hexatic-solid transition~\cite{anderson2017shape}, and $2,500$
hexagons for higher densities. In addition, we validated our analytical and
numerical approaches with systems of hard disks. In all cases, we initialized
systems into hexagonal lattices, thermalized and equilibrated at a selected density, and sampled
over $1\times 10^7$ Monte Carlo (MC) sweeps at high density, and $5\times 10^7$
MC sweeps at low density. Four independent samples were run at each density.
From these MC data we computed dynamical matrix elements at all wavevectors
$\vec k$ commensurate with the periodic boundary conditions using ensemble
averages of statistically independent samples.

To calculate dynamical matrix elements, we constructed a linear elastic model
(see SI for details) of the system that assumes that anisotropic hard core
interactions can be modeled via effective, entropic pair
interactions~\cite{entint}, $\mathcal U_{1,2}(\left|{\vec r_1-\vec
r_2}\right|,\theta_1,\theta_2)$, which yield the parameters
\begin{equation}
  \frac{\partial^2\mathcal U_{1,2}}{\partial r_{1,2}^2} = A, \quad
  \frac{\partial^2\mathcal U_{1,2}}{\partial(\theta_1+\theta_2)^2} = B_+, \quad
  \frac{\partial^2\mathcal U_{1,2}}{\partial(\theta_1-\theta_2)^2} = B_-,
\end{equation}
where $\vec r_i$ and $\theta_i$ describe particle positions and orientations.
$A$ and $B_\pm$ determine the eigenvalues of the dynamical matrix (see Fig.\
\ref{force_diagram} for illustration and SI for mathematical form) and thus
function as effective, entropic stiffness
parameters~\cite{effectivefrankconstant}. We extracted them from dispersion
relations computed with MC data using least squares fitting.

At high densities, when dislocations are absent, we use displacement covariance
analysis (DCA)~\cite{dca} to compute dispersion relations. DCA uses
equipartition to relate small, thermal deviations $x_i$ to a dynamical matrix
$K_{ij}$ according to $\langle x_i x_j \rangle = k_BT K_{ij}^{-1},$ where the
$x_i$ of interest are phonon modes $u_\mu(\vec k)$,
\begin{equation}
  u_\mu(\vec k) = \frac{1}{\sqrt{N}}\sum_\alpha (r^\alpha_\mu - \langle
  r^\alpha_\mu \rangle) e^{-i \vec k \cdot \langle \vec r^\alpha \rangle},
  \label{DCA_u_def}
\end{equation}
and libron modes $\theta(\vec k)$,
\begin{equation}
  \theta(\vec k) = \frac{1}{\sqrt{N}}\sum_\alpha
  \left[(\theta^\alpha-\langle\theta^\alpha\rangle) \, \mathrm{mod} \,
  \frac{2\pi}{6} \right] e^{-i \vec{k}\cdot \langle \vec r^\alpha \rangle}.
\end{equation}
Here, $r^\alpha_\mu$ is the $\mu$ spatial component of the position of the
center of particle $\alpha$, $\theta^\alpha$ is the orientation of the body of
the particle with respect to the positive $x$-axis, and $N$ is the total number
of particles. Particles' rotation symmetry is removed by the modulo $2\pi/6$
operation. We fix units by taking the particle circumdiameter and the thermal
energy $k_BT$ to be unity. At lower densities, where dislocations become
important and particles are able to diffuse from their lattice positions, we
adapt a method that decomposes density fluctuations into contributions from
transverse and longitudinal hydrodynamic wave modes~\cite{Walz2010}. This method
is tolerant of dislocations and does not assume particles vibrate closely around
well-defined lattice positions, but is more computationally intensive than
simple DCA. See SI for details.

We determined the density-dependent, effective elastic constants for both phonon
and libron modes in hard hexagon crystals. We validated our results for $A$ by
extracting the melting transition density, at which the
Kosterlitz-Thouless-Halperin-Nelson-Young (KTHNY) theory~\cite{Nelson1979}
predicts $2A/\sqrt{3}$ should change discontinuously from the universal value of
$16\pi$ to $0$, and find good agreement with melting densities reported in
\cite{anderson2017shape} (see Fig.\ \ref{couplings}). Results for hard disks
are also shown for comparison. For both hexagons and disks Fig.\
\ref{couplings} data are given at densities $\phi$ expressed with respect to
dense packing density $\phi_d$ (hexagons: 1, disks: $\pi/(2\sqrt{3})$) and melting
density $\phi_m$ (hexagons: $0.710$, disks: $0.719$ \cite{anderson2017shape}) in
the form $(\phi-\phi_m)/(\phi_d-\phi_m)$. The near collapse of the data
observed for the phonon elastic constant between hexagons and disks reveals the
interesting result that using ordinary, translational sound, it is difficult at moderate densities and above to
distinguish the difference between hexagon- and disk-shaped particles if density
is scaled appropriately.

We present in Fig.\ \ref{dispersion} phonon and libron dispersion relations for
a range of densities for high symmetry points in the Brillouin zone. These
frequencies are computed from eigenvalues of the dynamical matrix multiplied by
the inverse mass matrix, which has diagonal elements $1/m$ for phonon modes
and $1/I = 48/(5m)$ in reduced units for libron modes. We find several features of
the dispersion relations that are common at all densities. First, we find
deviations between our simulation data and the linear elastic model are only
appreciable near the edge of the Brillouin zone. The edges correspond to short
distances, and is precisely where we would expect a linear elastic model of hard
interactions to break down. Although deviations increase with density as
expected, we find that the linear elastic model provides a good approximation
for most of the Brillouin zone, even for relatively high densities of 0.9.
Second, we find that phonon modes are always ungapped, which is guaranteed by
Goldstone's theorem. Third, we find that the libron modes are gapped at all
densities, but that their group velocity increases rapidly with system density.

What do these features imply the contributions are to the macroscopic entropy of
the system? It can be shown (see SI) that the fluctuation spectrum yields a
mode-by-mode decomposition of the configurational entropy (up to overall
additive constants) according to
\begin{equation}
  S = -\frac{1}{2}\sum_{i,\vec{k}}\ln(\omega_i^2({\vec{k}}))
  \; . \label{entropy}
\end{equation}
Here, $\omega_i(\vec{k})$ are the dispersion frequencies, and the sum extends
over all mode types $i$ corresponding to longitudinal and transverse phonon
modes and libron modes at all wavevectors $\vec k$ in the first Brillouin zone.
Brillouin zones are shown in Fig.\ \ref{dispersion}, as ``entropy maps,'' in
which all points are colored according to the mode that provides the dominant
contribution to the system entropy at that wavevector, using Eq.\ \ref{entropy}.

This analysis produces a similar entropy map at densities just above the
hexatic-solid transition (0.722) and close to dense packing (0.9) in which
entropy at all wavevectors comes predominantly from translational disorder. The
predominance of translational disorder just above the hexatic-solid transition
arises because although the libron group velocity is small, the gap in the
libron spectrum is sufficiently large to put the libron dispersion above the
phonon dispersion throughout the Brillouin zone (save for wavevectors at the
very edge of the Brillouin zone where we expect the harmonic approximation to
breakdown). This is in contrast to what occurs at high densities. At high
densities the phonon group velocity itself has increased by an order of
magnitude (cf.\ vertical scales in Fig.\ \ref{dispersion}a vs.\ d). In
addition, we find there is an increase in the libron gap, but that this increase
is much less dramatic than the increase in the libron group velocity, which
becomes similar in scale to the phonon group velocity.

Our observation of ungapped phonon modes is a consequence of Goldstone's
theorem, but our observation of a gapped libron spectrum is the result of more
subtle field-theoretic effects. For systems of isotropic particles, it has been
shown that although crystals break both translational and rotational symmetry,
the absence of Goldstone modes associated with rotational symmetry breaking is
due to linear relations among translation and rotation generators that render
the Goldstone modes associated with rotational symmetry breaking
redundant~\cite{Watanabe2013}. For systems of anisotropic particles, rotation
generators contain additional terms that correspond to particle body rotations
which introduce additional degrees of freedom. These degrees of freedom can be
shown (see SI) to always couple to the global rotation generator as a mass term
in a linear elastic model. The existence of coupling between the generator of
global rotations and the generator of body rotations implies that, in generic
situations, the libron modes for anisotropic colloids or molecules should be
gapped, whereas Goldstone's theorem requires that phonon modes remain
ungapped~\cite{inverseHiggs}. Thus, librational gaps should appear in any
crystal or hexatic phase of anisotropic particles with global orientational
order. Because of this, we expect that systems of anisotropic particles that
melt from crystals with quasi-long range translational and long range orientational order into
hexatic or other phases with quasi-long range orientational order but without
translational order will exhibit entropy maps similar to those found here.  We
note that, practically, the identification of the signatures of solid-hexatic
transition found here were far less computationally intensive to produce than
approaches that were used to confirm the transition
elsewhere~\cite{anderson2017shape}. In contrast, if a system were to ``melt''
from a crystal with well-defined orientational order to a plastic crystal, we
would expect a dramatically different entropy map.

At high densities, the predominance of translational disorder at all wavevectors
arose through a combination of the gap in the libron spectrum and the comparable
group velocity for phonon and libron modes. In our model systems we found that
translational disorder predominated at all wavevectors at all densities at or
above 0.9. Because the libron gap is a consequence of global orientational
order, and because dense packing of highly anisotropic shapes generically leads
to strong constraints on both particle positions and orientations, we would expect
that similar computations in other systems would lead to similar phonon and
libron group velocities. We therefore expect the predominance of translational
disorder to be a generic feature of densely packed colloids. If this is so, it
would help to explain recently reported, unexpected results from inverse-design
\cite{digitalalchemy} indicating that non-space-filling shapes are
thermodynamically preferred in model systems of hard colloids even at densities 
above 0.99.\cite{packingassembly} In the context of the present results, this
finding can be rationalized because systems stand to gain more entropy through rotational modes -- which the entropy map shows are restricted at all wavevectors at high densities --  than through translational modes.  This is most easily achieved by slight rounding of the particles into non-space-filling shapes, as observed in Ref. ~\cite{packingassembly}.

We decomposed the total entropy of an entropically stabilized solid into
microscopic contributions from collective modes of particle motion. We showed
for a simple hard hexagon solid that translational and rotational degrees of
freedom contribute different amounts of entropy at different wavevectors,
extending the Onsager picture of hard rod liquid crystals~\cite{Onsager1949,
freons, frenkel} to hard particle crystals. Our method is applicable to any
entropically stabilized solid in which the particles exhibit harmonic motion
about equilibrium positions, and it is generalizable to three dimensions.

\acknowledgments
We thank X.\ Mao and R.\ Kamien for helpful discussions. This work was
supported by the National Science Foundation Graduate Research Fellowship Grant
No.\ DGE 1256260 to J.A., and a Simons Foundation Investigator Award to S.C.G. 
Any opinions, findings, and conclusions or recommendations
expressed in this material are those of the authors and do not necessarily
reflect the views of the National Science Foundation. Data analysis in this work
was performed partially using Freud, an open-source Python-driven analysis
toolkit~\cite{freudsoftware}. The computational workflow and data management
were primarily supported by the signac data management framework~\cite{signac}.

\end{document}